\documentclass[a4paper,11pt]{article}
\usepackage{pos}
\renewcommand{\logo}{\relax}

\title{The SNO+ Experiment: Reactor \& Solar $\nu$ Prospects}

\manuallySeparateAuthors
\author*{Benjamin Tam}
\author{ for the SNO+ collaboration}

\affiliation{Department of Physics, Engineering Physics \& Astronomy, Queen's University,\\
  64 Bader Ln, Kingston, Canada}

\emailAdd{benjamin.tam@queensu.ca}

\abstract{

The SNO+ experiment is a large-scale, multipurpose neutrino experiment situated 2\,km underground at SNOLAB in Canada. Successor to the Sudbury Neutrino Observatory, the SNO+ detector has inherited much of the original infrastructure including the 12-m diameter acrylic vessel which serves as the main detector body. Initially filled with ultrapure water, the SNO+ experiment has completed operations as a water Cherenkov detector, having set new limits on multiple invisible nucleon decay modes, performed measurements on $^8$B solar neutrinos, and made the first observation of reactor antineutrinos in pure water. The detector medium has now been replaced with liquid scintillator, and a new physics programme is being pursued including measurements of solar neutrinos and $\Delta$m$^2_{12}$ from reactor antineutrinos. The liquid scintillator will be doped with >4 tonnes of $^{\mathrm{nat}}$Te to enable a search for neutrinoless double beta decay.

}

\FullConference{%
  Neutrino Oscillation Workshop-NOW2022\\
  4-11 September, 2022\\
  Rosa Marina (Ostuni, Italy)}

\begin{document}
\maketitle

\section{The SNO+ Experiment}

The SNO+ experiment is a low background kilotonne-scale neutrino experiment located at the SNOLAB facility in Canada. The successor to the Sudbury Neutrino Observatory (SNO) \cite{sno}, the SNO+ detector has inherited much of the infrastructure, with most aspects upgraded and refurbished to accommodate an expanded physics programme utilising liquid scintillator \cite{detector}.

The main body of the SNO+ experiment is a 6-m radius acrylic vessel (AV) with a 905\,m$^3$ volume. Events within the AV are observed with over 9000 inward-facing 8" photomultiplier tubes (PMTs). Mounted on an 8.9\,m radius steel PMT support structure (PSUP) that encompasses the AV, each PMT is attached with a 27-cm diameter concentrator providing a 54\% effective photocoverage. The PMTs are calibrated using a ropes-based source deployment system, as well as an LED/laser calibration system that is mounted on the PSUP \cite{optics}.
To mitigate external backgrounds, the detector has a N$_2$ cover gas system that blankets the AV, is submerged in 7000\,m$^3$ of water shielding, and is situated under a 2070\,m rock overburden, corresponding to $6010\,\mathrm{m.w.e.}$ and a muon rate of 0.286$\pm$0.009\,$\mu$/m$^2$/d \cite{muon}. A complete description of the SNO+ detector and all hardware aspects of the experiment is published in \cite{detector}.

SNO+ operations are separated into three phases distinguished by the target medium deployed in the AV.  During the ``water phase'', the detector was filled with 905\,tonnes of ultrapure water and operated as a water Cherenkov detector from May 2017 - July 2019. The water was then replaced with 780\,tonnes of liquid scintillator, increasing the light yield of the detector by a factor of $\sim$50 thereby allowing for the study of lower energy processes at higher resolution. This ``scintillator phase'' is ongoing, having started in April 2022. In the upcoming ``tellurium phase'', the scintillator will be doped with >4\,tonnes of $^{\mathrm{nat}}$Te, enabling a search for neutrinoless double beta decay ($0\nu\beta\beta$).

The physics goals of the water phase included a search for invisible nucleon decay, solar neutrinos, and reactor antineutrinos. In the scintillator phase, these goals are expanded to also include geoneutrinos, supernova neutrinos, and a search for dark matter candidates and axion-like particles. The tellurium phase will feature a search for $0\nu\beta\beta$ in addition to continuing the scintillator phase physics programme. 

\section{Water Phase Results}
\label{sec:water}

The SNO+ water phase was made up of two datasets. The first set occurred during detector commissioning and lasted for 115 live days. The second set occurred after the detector was fully commissioned, and lasted for 190 live days. In particular, a N$_2$ cover gas system was introduced to protect the AV from air\footnote{The air at SNOLAB has a $^{222}$Rn rate of $\sim$100 Bq/m$^3$.}, thereby allowing the second dataset to feature lower background analyses.

\paragraph{Invisible Nucleon Decay}

Invisible nucleon decay (IND) is a process where free protons or neutrons decay to modes where no visible energy is deposited during the decay itself. Since this would violate baryon number conservation, an observation of IND would provide direct evidence of physics Beyond the Standard Model. Many experiments have searched for IND over the last several decades, considering both mononucleon ($p$, $n$) and dinucleon ($pp$, $nn$, $pn$) modes. An IND search using the first SNO+ water phase period set new limits on the $p$, $pp$, and $np$ modes \cite{ind1}. Recently, an improved search utilising the second dataset not only improved upon those limits, but also set a new limit on the $n$ decay mode. The new analysis and limits are published in \cite{ind2}.


\paragraph{Reactor Antineutrinos}
60\% of SNO+ antineutrino events occur from 3 reactors (18 cores) situated 240\,km, 340\,km, and 350\,km away from the detector. Reactor antineutrinos are detected in SNO+ using inverse beta decays (IBDs) on protons: $\bar{\nu} + p \rightarrow e^+ + n$. IBDs are identified when a prompt signal from the $e^+$ is in coincidence with a delayed 2.2\,MeV $\gamma$ produced when the $n$ thermalises before being captured by a nucleus. As detecting this $\gamma$ requires low thresholds and ways to deal with backgrounds below 3\,MeV, IBDs have not been previously observed in pure water Cherenkov detectors. However, SNO+ has achieved the lowest energy threshold of any large Cherenkov detector (1.4 MeV/$e$) and has a neutron detection efficiency of 50\% \cite{neutron}. From these advantages, SNO+ has made the first observation of reactor antineutrinos in pure water with a significance of 3.5$\mathrm{\sigma}$ \cite{antinu}.

\begin{figure}[h]
\centering
\includegraphics[width=14cm]{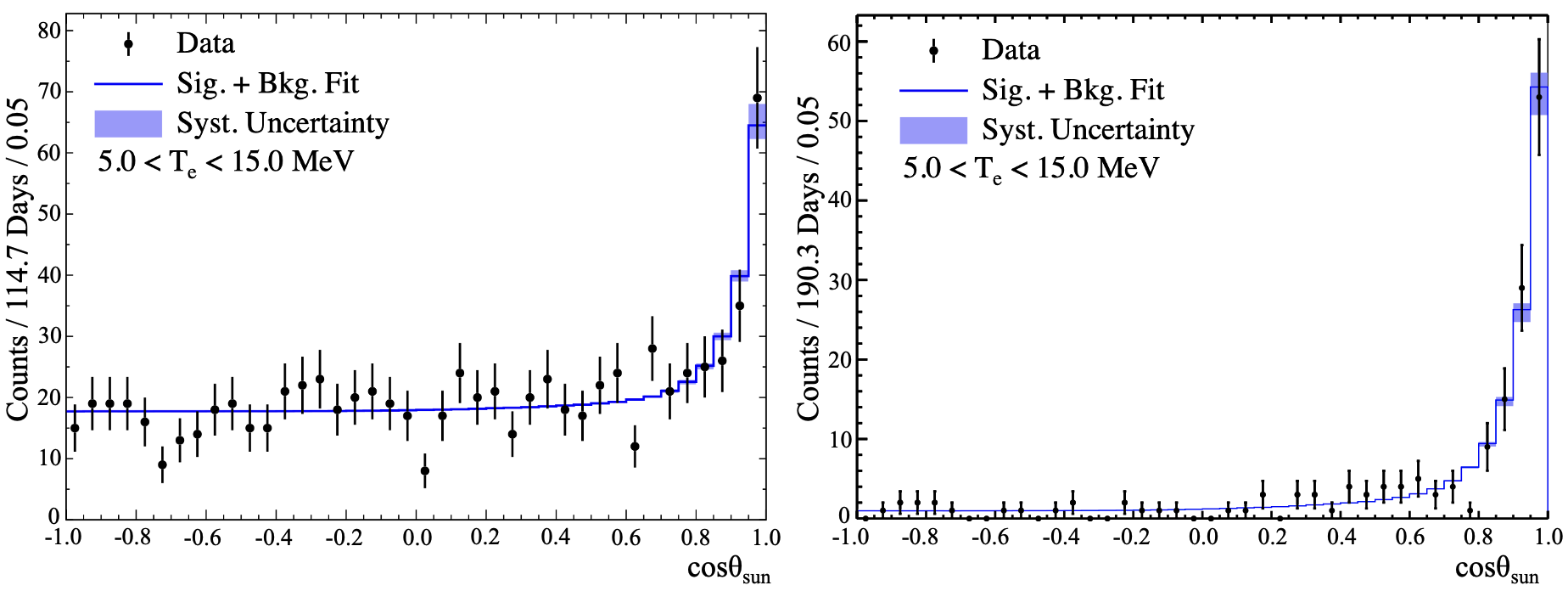}
\caption{Distribution of event direction with respect to the solar direction using (left) the first dataset as published in \cite{solar} and (right) the second dataset. The backgrounds were improved following the full commissioning of the detector.}
\label{fig:solar}
\end{figure}

\paragraph{Solar Neutrinos}
The identification of solar neutrinos in the SNO+ water phase utilises the angle between the direction of the electron recoil and the position of the sun,  cos$\theta_\mathrm{sun}$. Therefore, no \textit{a priori} knowledge of the backgrounds are required for a robust flux measurement. The $^8$B solar neutrino flux was initially measured with the first dataset to be $\Phi=5.95^{+0.75}_{-0.71}\mathrm{(stat.)}^{+0.28}_{-0.30}\mathrm{(syst.)}$, consistent with measurements by SNO and Super Kamiokande \cite{solar}. Following the full commissioning of the detector, the backgrounds have been further suppressed in the second dataset, as seen in Figure \ref{fig:solar}. The measurement using the improved second dataset has been completed, and will be shown in an upcoming publication.

\section{Scintillator Phase Prospects}

The SNO+ collaboration developed a new liquid scintillator based on linear alkylbenzene (LAB), as existing widespread liquid scintillators such as pseudocumene (PC) and phenyl-o-xylylethane (PXE) are not compatible with the acrylic that makes up the AV \cite{scint}. LAB-based liquid scintillators have since been successfully used in a number of other large scale experiments. In SNO+, LAB is doped with the fluor diphenyloxazole (PPO) at a concentration of $2.2\,\mathrm{g_{PPO}/L_{LAB}}$ in order to emit in the $\sim$390\,nm range. Prior to deployment within the AV, the scintillator was purified through multi-stage distillation and N$_2$ stripping using a purpose-built purification plant. Following deployment, the entire volume was further recirculated through the plant over 5 times. The purity of the scintillator was verified through an extensive suite of hourly measurements during both filling and recirculation of the detector. The intrinsic backgrounds in the scintillator have been measured to be $(4.7\pm1.2)\times 10^{-17}$\,gU/g and $(5.3\pm1.5)\times10^{-17}$\,gTh/g. Scintillator phase data taking began in April 2022 and is ongoing.

\paragraph{Reactor Antineutrinos}
As mentioned in section \ref{sec:water}, most antineutrino events in SNO+ are produced in 3 reactors 240-350\,km away. Coupled with the higher light yield from the replacement of water with scintillator, SNO+ has an excellent L/E for sensitivity to $\Delta \mathrm{m}^2_{21}$. This is particularly important towards resolving the current 1.5$\sigma$ tension between reactor (KamLAND) and solar (SNO and Super Kamiokande) measurements. The most significant background towards this analysis are when $\alpha$ decays from $^{210}$Po in the detector cause $^{13}$C($\alpha,n$)$^{16}$O events: $\alpha+\mathrm{^{13}C}\rightarrow \mathrm{^{16}O}+n$. These $(\alpha,n)$ events are challenging to distinguish from IBD events. However, SNO+ has developed a pulse shape discriminator that utilises the slightly different timing profile of proton recoils from $(\alpha,n)$ events, and is now capable of rejecting these backgrounds by a factor of 10.

\paragraph{Solar Neutrinos}
With the low backgrounds and use of liquid scintillator, SNO+ will measure the $^8$B solar neutrino-electron recoil energy spectrum as low as 3\,MeV or lower. Deviations in the spectrum from the expected Mikheyev, Smirnov, and Wolfenstein (MSW) behaviour will be used to probe for new physics such as Non Standard Interactions \cite{nsi}.
Traditionally, the isotropic nature of scintillation makes event directionality challenging to determine, rendering the cos$\theta_\mathrm{sun}$ technique used to measure solar neutrino fluxes in the water phase ineffective. However, by fitting prompt timing profiles to combined Cherenkov-scintillation 2D distributions, SNO+ has developed a new event-by-event direction reconstruction technique - the first in a liquid scintillation experiment. This capability will be demonstrated in a forthcoming publication.

\section{Future Prospects}

Following the scintillator phase, the SNO+ liquid scintillator will be loaded with $^\mathrm{130}$Te, a $0\nu\beta\beta$ candidate isotope. This will be done using a novel metal-loading technique developed by the SNO+ collaboration. Telluric acid (TeA) is diolised to form tellurium butanediol (TeBD), which readily dissolves in LAB. N,N-dimethyldodecylamine (DDA) and 1,4-bis(2-methylstyrl)benzene (bis-MSB) will also be added to boost stability and light yield, respectively. The TeA, which has been stored underground since 2015 to minimise cosmogenic backgrounds, will be purified, diolised, and deployed using multiple purpose-built chemical plants. The initial loading will be 0.5\% $^{\mathrm{nat}}$Te\footnote{$^{130}$Te has a natural abundance of 34\%} by mass, providing a $0\nu\beta\beta$ sensitivity of $\mathrm{T}^{0\nu}_{1/2}=2\times10^{26}$\,yr after 3 years of data. Further loading is possible and planned; this would improve the sensitivity and may probe below inverted ordering parameter space \cite{biller}. 

The deployment of Te in SNO+ is planned to start in 2024. In the tellurium phase, the ongoing scintillator phase physics programme will be continued in addition to the search for $0\nu\beta\beta$.

\acknowledgments
This work is supported by ASRIP, CIFAR, CFI, DF, DOE, ERC, FCT, FedNor, NSERC, NSF, Ontario MRI, Queen’s University, STFC, and UC Berkeley, and have benefited from services provided by EGI, GridPP and Compute Canada. Benjamin Tam is funded by NSERC (Natural Sciences and Engineering Research Council of Canada), Queen's University, and the Walter C. Sumner Foundation. Vale and SNOLAB is thanked for their valuable support.

\end{document}